\documentclass[letterpaper]{article} 
\usepackage{aaai24}  
\usepackage{times}  
\usepackage{helvet}  
\usepackage{courier}  
\usepackage[hyphens]{url}  
\usepackage{graphicx} 
\usepackage{natbib}  
\usepackage{caption} 
\usepackage{algorithm}
\usepackage{algorithmic}

\usepackage{newfloat}
\usepackage{listings}
\DeclareCaptionStyle{ruled}{labelfont=normalfont,labelsep=colon,strut=off} 
\floatstyle{ruled}
\newfloat{listing}{tb}{lst}{}
\floatname{listing}{Listing}
\title{How Can Large Language Models Enable Better Socially Assistive\\ Human-Robot Interaction: A Brief Survey}
\author{
    Zhonghao Shi, Ellen Landrum,
    Amy O'Connell,
    Mina Kian,
    Leticia Pinto-Alva,
    Kaleen Shrestha,
    Xiaoyuan Zhu,
    Maja J Matari\'c
}
\affiliations{
    University of Southern California, Los Angeles, United States\\

     \{zhonghas, elandrum, amy.dell, kian, pintoalv, kshresth, xzhu9839, mataric\}@usc.edu
}

\usepackage{bibentry}

\begin{document}

\maketitle

\begin{abstract}
Socially assistive robots (SARs) have shown great success in providing personalized cognitive-affective support for user populations with special needs such as older adults, children with autism spectrum disorder (ASD), and individuals with mental health challenges. The large body of work on SAR demonstrates its potential to provide at-home support that complements clinic-based interventions delivered by mental health professionals, making these interventions more effective and accessible. However, there are still several major technical challenges that hinder SAR-mediated interactions and interventions from reaching human-level social intelligence and efficacy. With the recent advances in large language models (LLMs), there is an increased potential for novel applications within the field of SAR that can significantly expand the current capabilities of SARs. However, incorporating LLMs  introduces new risks and ethical concerns that have not yet been encountered, and must be carefully be addressed to safely deploy these more advanced systems.
In this work, we aim to conduct a brief survey on the use of LLMs in SAR technologies, and discuss the potentials and risks of applying LLMs to the following three major technical challenges of SAR: 1) natural language dialog; 2) multimodal understanding; 3) LLMs as robot policies. 
\end{abstract}

\section{Introduction}

At the intersection of assistive robots and socially interactive robots, {\it socially assistive robots} aim to provide assistance to support users' cognitive-affective needs through social interaction~\cite{feil2011socially}. Past studies have demonstrated the benefits and potential of deploying SARs to support a diverse set of user populations, including individuals with mental health challenges~\cite{scoglio2019use}, children with ASD~\cite{cabibihan2013robots}, older adults~\cite{shishehgar2018systematic}, and K-12 students~\cite{randall2019survey}. With more affordable robot hardware~\cite{koh2021impacts, pinto2022physical, koh2022usability}, SARs have the potential to lower the socio-economic barriers that limit access to personalized therapies, companionship, and education. 
However, prior work has also demonstrated that the existing SAR interactions have not yet been able to understand multimodal (language and visual) social dynamics~\cite{robinson2023robotic, li2021intention} and respond with human-like dialog~\cite{skantze2021turn} and actions~\cite{akalin2021reinforcement}.

With the recent advances in natural language processing (NLP) research, large language models (LLMs) such as GPT-4 have shown tremendous success in tasks both within the field of NLP (such as language modeling, question answering, and translation)~\cite{achiam2023gpt} and outside (such as programming~\cite{xu2022systematic}, robot planning~\cite{singh2023progprompt}, and autonomous driving~\cite{cui2024survey}). These capabilities of LLMs may open new possibilities for tackling the core technical challenges of SAR, and help us get closer to achieving more effective, human-level social assistance for users with differences.

In this work, we categorize the core technical challenges of SAR into three areas: 1) natural language dialog; 2) multimodal user understanding; 3) LLMs as robot policies. To survey the existing work on LLM-powered SARs in these three categories of technical challenges, we used Google Scholar to search the relevant papers in major human-robot interaction conferences, journals, and arXiv. In each of the following sections, we aim to: identify the potential of LLMs for SAR, survey existing work, and discuss future directions. This research was supported by the National Science Foundation Grant ITE-2236320 and IIS-1925083.

\section{Natural Language Dialogue }

Natural language dialogue is at the core of human-centered social interaction. Yet, prior to the recent breakthroughs in LLM technologies, SARs mainly relied on Wizard-of-Oz teleoperation~\cite{erich2017systematic} or predefined rule-based dialogue management systems~\cite{erich2017systematic, youssef2022survey}. SARs employing traditional non-LLM-based conversational systems are limited by their inability to accurately interpret human dialogue, limited vocabulary in dialogue generation, lack of understanding of context and ability to personalize, and lack of ability to effectively utilize online resources~\cite{grassi2022knowledge}.




By applying state-of-art LLM models such as GPT-4~\cite{achiam2023gpt}, recent work on LLM-powered SAR has been mainly focused on enabling more accurate dialog understanding and more human-like and context-aware dialogue generation. LLM-powered SARs are able to produce varied dialogue while staying on topic ~\cite{billing2023language}. They can engage in more natural, flexible conversation with users from populations of interest, such as older adults and children with autism spectrum disorder~\cite{bertacchini2023social}. \citet{spitale2023vita} designed an LLM-powered SAR as a motivational coach with both informative and emotional objectives, demonstrating that LLMs can be used to understand long-horizon context and enable long-term personalization. In instructional settings, LLM-powered SARs combine the vast knowledge base and interactive content-delivery of LLMs with the the capabilities and engaging nature of physically embodied agents~\cite{wake2023gpt}. 

Despite this progress, \citet{irfan2023between} showed that hallucinations, obsolete information, latency, and disengagement cues may still cause user frustration and confusion, which could be detrimental to socially assistive human-robot interaction. 
More exploration is needed to overcome these limitations and harness the power of LLMs in ways that better align with the goals of SAR interactions.

\section{Multimodal User Understanding}

To enable successful socially assistive human-robot interactions, a SAR needs to understand the user's cognitive-affective state (user engagement, affect, and intent) from multimodal perceptual data (language, visual, and audio)~\cite{youssef2022survey}. Existing work on multimodal social understanding has relied on training and fine-tuning machine learning (ML) models with data collected from previous SAR deployments~\cite{robinson2023robotic}. However, the definition of cognitive-affective states may vary in different social contexts. Due to the independent and identically distributed (IID) assumption made by ML model training~\cite{wang2022generalizing}, existing ML models struggle to generalize effectively and quickly to test data that are distributed differently from the training data, particularly in the context of SAR~\cite{shi2022toward}.

Multimodal language models (MLMs) such as state-of-the-art vision-language models like CLIP~\cite{radford2021learning}, ALIGN~\cite{jia2021scaling}, and GPT-4V~\cite{achiam2023gpt}, have shown promising zero-shot performance on a variety of human-centered visual tasks~\cite{zhang2023vision, wu2023gpt4vis}. Furthermore, these MLMs also demonstrate impressive few-shot capabilities of quickly adapting via prompting with natural language~\cite{ge2023mllm}. This indicates that MLMs may also be capable of adapting to novel social context for more generalizable and accurate multimodal social understanding. Despite the recent progress in computer vision and robotics, using MLMs for SAR is still largely unexplored, but this direction of research shows great potential for significantly enabling better multimodal social understanding for socially assistive human-robot interaction.

\section{LLMs as Robot Policies}

In an ideal socially assistive human-robot interaction, a SAR should fluently learn and reason about the user's states and provide the best feedback or action as assistance. The space of user's states and robot feedback or actions can be large and continuous~\cite{clabaugh2019escaping} and existing approaches to encoding SAR policies, such as rule-based system and reinforcement learning, are not efficiently trained or sufficiently robust on large and continuous spaces with limited amounts of data~\cite{akalin2021reinforcement}. Past work has often circumvented this problem by constraining interactions to pre-defined tablet/computer games with small user state and action spaces~\cite{clabaugh2019escaping}. LLMs may help to relax this constraint and enable more natural interaction by allowing continuous, more human-like formulation of space for user states and robot actions.

Recent studies in robotics and NLP have successfully employed LLMs as robot policies in the setting including autonomous driving~\cite{cui2024survey}, robot task planning~\cite{singh2023progprompt, saycan-ahn2022i}, social common sense~\cite{sap-etal-2019-social} and social reasoning~\cite{gandhi2023understanding}. In the field of SAR, existing work has mainly focused on applying LLMs to matching the affective state of robot feedback with the sentiment of the robot's dialog~\cite{lee2023developing, mishra2023real, lim2023sign}. These studies have shown that LLM-powered robot policies for gesture matching enable more context-aware and natural robot gesture. Research using LLMs as robot policies has not yet explored 1) how to enable SARs to form robot policies for spontaneous tasks instead of only pre-defined ones, such as helping children with ASD navigate through an unpleasant social interaction they just encountered during their school day; 2) how to enable SAR policies to engage users with educational tasks while keeping them both challenged and encouraged; 3) how to reason socially about the user's intent and needs with partially observable information based on multimodal data; and 4) how to enable personalized SAR policies to quickly align with each user's unique needs, personality, and values, so the SAR can be more helpful and empathetic to each user.





\section{Risks and Safety Considerations}

Because SAR aims to support vulnerable populations including children and individuals with physical and/or mental health challenges~\cite{mataric2016socially}, it is crucial to ensure absolute safety during socially assistive human-robot interactions. Despite the great success in LLMs, their lack of explainability and theoretical safety guarantees~\cite{huang2023survey} may introduce significant risks and concerns by 1) amplifying unfairness and human bias~\cite{acerbi2023large}; 2) harming data security and privacy by unethical use of personal data~\cite{liyanage2020ethical}; and 3) hallucination behaviors causing potential harms to users~\cite{zhang2023siren}. For these reasons, the trustworthiness of LLM-powered SAR needs to be extensively evaluated before autonomous system can be deployed with vulnerable populations without human monitoring and oversight. As the first survey paper on this topic, this work aims to inform and stimulate research toward leveraging the tremendous potential of LLM-powered SARs and address the risks and safety concerns.

\bibliography{aaai24}

\end{document}